\documentclass[aps,prl,twocolumn,superscriptaddress]{revtex4}
\usepackage{graphicx}
\usepackage{amsmath,amssymb}
\usepackage{bm}
\usepackage{verbatim}
\usepackage{ulem}
\usepackage{color}

\newcommand{\V}[1]{{\bm #1}}

\newcommand{\rmb}{{\rm b}}

\begin{document}
\title{Photoemission spectra of massless Dirac fermions on the verge of
exciton condensation}
\author{Stefan Rist}
\affiliation{CNR-SPIN, Corso Perrone 24, I-16152 Genova and NEST, Scuola Normale
Superiore, I-56126 Pisa, Italy}
\author{A.A. Varlamov}
\affiliation{CNR-SPIN, Tor Vergata, Viale del Politecnico 1, I-00133 Rome, Italy}
\author{A.H. MacDonald}
\affiliation{Department of Physics, University of Texas at Austin, Austin, Texas 78712,
USA}
\author{Rosario Fazio}
\affiliation{NEST, Scuola Normale Superiore and Istituto di Nanoscienze-CNR, I-56126
Pisa, Italy}
\author{Marco Polini}
\affiliation{NEST, Istituto Nanoscienze-CNR and Scuola Normale Superiore, I-56126 Pisa,
Italy}

\begin{abstract}
Angle-resolved photoemission spectroscopy (ARPES) is a powerful probe of
electron correlations in two-dimensional layered materials. In this Letter
we demonstrate that ARPES can be used to probe the onset of exciton
condensation in spatially-separated systems of electrons and holes created
by gating techniques in either double-layer graphene or topological-insulator thin
films.
\end{abstract}

\maketitle

\textit{Introduction---}Electron-hole pairs can undergo Bose-Einstein
condensation forming a macroscopically coherent neutral fluid, the exciton
condensate (EC)~\cite{Blatt_pr_1962,keldysh_jept_1968,Brandt_jept_1972,Abrikosov_jept_1973}. 
EC states have been predicted theoretically in a variety of different systems.  Spatially separated
electrons and holes located on two nearby layers (like those that can be
created in GaAs/AlGaAs semiconductor double quantum wells) were considered first
some time ago~\cite{ lozovik_jept_1975}.  These double-layer exciton condensates
(DLECs) possess spontaneous inter-layer coherence and support counter-flow 
supercurrents.  The DLEC has
been discussed more recently in the context of quantum Hall bilayers at total filling factor $\nu_{\rm T}=1$~\cite{Fertig,Wen,eisenstein_macdonald_nature_2004,su_naturephys_2008}. The first
clear experimental evidence for exciton condensation was obtained
a decade ago in 
quantum Hall bilayer DLECs~\cite{spielman_prl_2000}, long after the state was predicted. 

There is an ongoing effort to find new systems which display exciton
condensation.  In addition to the fundamental interest in understanding when this
type of order occurs in nature, there is a practical interest in finding
ECs that occur under less extreme physical conditions, in particular 
in the absence of strong magnetic fields and, possibly, at non-cryogenic
temperatures. Very recently exciton condensation has been predicted to occur
in double-layer graphene~\cite
{min_prb_2008,joglekar_prb_2008,lozovik_jept_2008, mink_prb_2011} and in
three-dimensional (3D) gated topological insulators (TIs)~\cite{franz}. 
In the TI case spontaneous coherence is established between topologically 
protected single-particle states on top and bottom surfaces.  Both double-layer graphene and TI thin-film 
systems are described at low energies by two massless Dirac fermion (MDF)
Hamiltonians which interact.   Since MDF systems are gapless and
truly two-dimensional (2D), the field-driven carrier densities that can be
achieved are much larger than in the semiconductor case. Weaker dielectric
screening and linearly dispersive conduction and valence bands help to
increase both interaction and disorder energy scales. Finally, MDF bands are
perfectly particle-hole symmetric, guaranteeing the almost complete nesting between conduction- and valence-band Fermi surfaces, which favors the coherent state.

\begin{center}
\begin{figure}[t]
\includegraphics[width = 1.0 \linewidth]{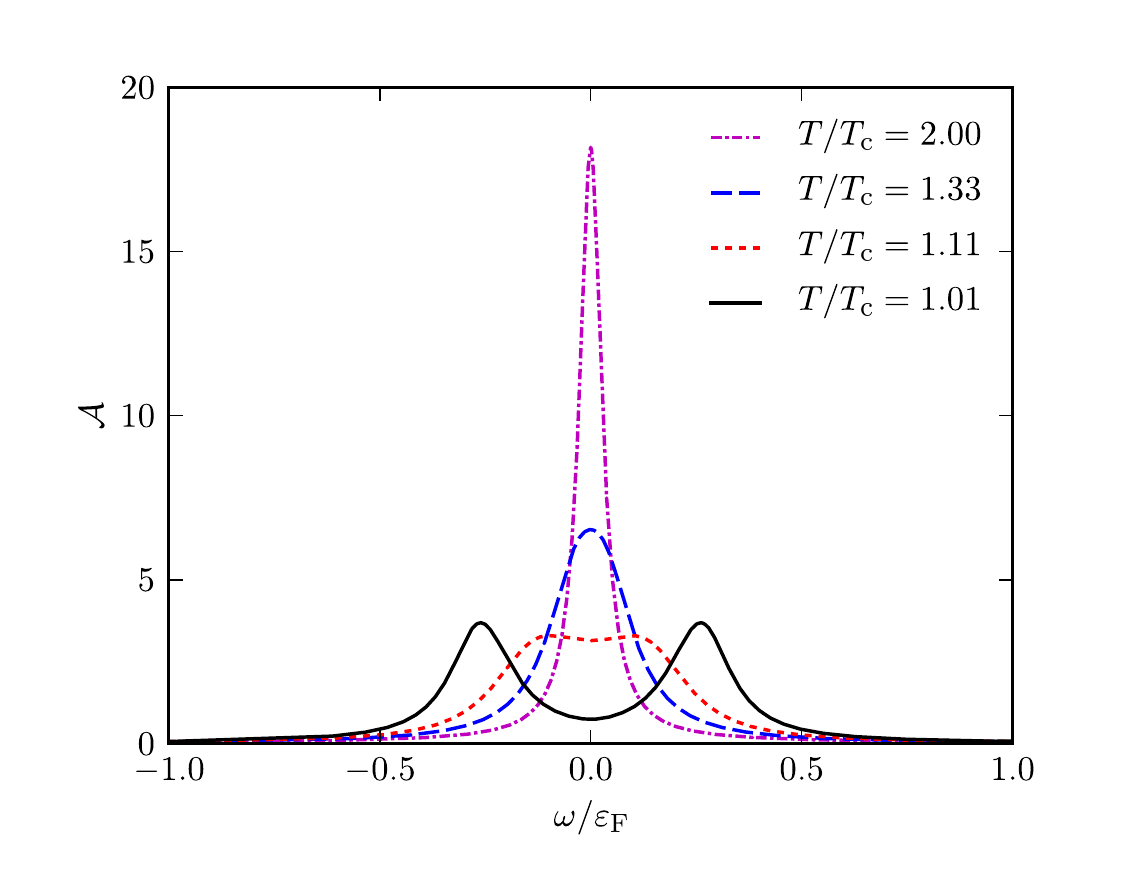}
\caption{(Color online) The spectral function of a massless-Dirac-fermion
double-layer system on the verge of exciton condensation. The spectral function $\mathcal{A}({\bm k},\protect\omega)$ (in units of $1/
\protect\varepsilon_{\mathrm{F}}$, where $\protect\varepsilon_{\mathrm{F}}$
is the Fermi energy) evaluated at $|{\bm k}| = k_{\mathrm{F}}$ is plotted as
a function of $\protect\omega/\protect\varepsilon_{\mathrm{F}}$ for
different values of $T_{\mathrm{c}}/T$. The single peak at the Fermi energy
in the high-temperature limit splits into two peaks as $T \to T_{\mathrm{c}}$
from above. All curves have been evaluated by setting $T = 0.1~T_{\mathrm{F}
} $.\label{fig:one}}
\end{figure}
\end{center}

Graphene double-layer systems can be realized by separating two graphene
layers by a dielectric~\cite{kim_prb_2011} (such as $\mathrm{Al}_2\mathrm{O}_3$) or by a few layers of a one-atom-thick insulator such as h-BN~\cite{dean_naturenano_2010,ponomarenko_naturephys_2011,britnell_science_2012,gorbachev_arXiv_2012}. 
In both cases inter-layer hybridization is negligible and the nearby
graphene layers are, from the point of view of single-particle physics,
isolated. Topologically-protected massless Dirac 2D electron systems
(MD2DESs) appear automatically~\cite{TIreviews} at the top and bottom
surfaces of a 3D TI thin film. 3D TIs in a slab geometry offer two surface
states that can be far enough apart to make single-electron tunneling
negligible, but close enough for Coulomb interactions between surfaces to be
important.

We report on an analysis undertaken in support of the experimental quest for
ECs with MDF character.  Because of the potential value of probes which 
can detect incipient order while still above
the critical temperature $T_{\rm c}$, we examine the influence of 
pairing fluctuations on the ARPES spectra that are commonly
used to probe two-dimensional systems like graphene.  Fluctuation corrections to
thermodynamic and transport observables have previously been extensively studied in
superconductors~\cite{varlamov}. General considerations based on the
uncertainty principle allow one to conclude that, like Cooper pairs in superconductors, excitons 
have a finite lifetime above $T_{\rm c}$
which is inversely proportional to the distance from the critical
temperature.  Exciton fluctuations will therefore influence the 
behavior of the double layer providing distinct signatures in the vicinity
of the EC instability. Mink \textit{et al.}~\cite{mink_prl_2012}, for
example, have investigated the role of fluctuations in a Coulomb-drag
transport setup, demonstrating that the drag resistivity grows logarithmically upon
lowering the temperature towards $T_{\mathrm{c}}$.

Experimental evidence for ECs in double-layer systems has so far been
largely based on transport measurements.  In the case of quantum
Hall systems the transport phenomenology~\cite{eisenstein_macdonald_nature_2004,su_naturephys_2008,spielman_prl_2000,nandi_nature_2012}
is spectacular in separately contacted coherent double layers. In this Letter we study the
effect of fluctuations on a contactless observable: the one-body spectral
function, which can be accessed by ARPES. ARPES, a photon-in electron-out experiment~\cite{damascelli_rmp_2003}, has been established as an extremely important tool
in studying the quasiparticle dynamics in graphene~\cite{ARPES-graphene} and
provided the first direct evidence for conducting surface states in 3D TIs~\cite{ARPES-TI}.

\begin{figure}[tbp]
\includegraphics[width = 1.0 \linewidth]{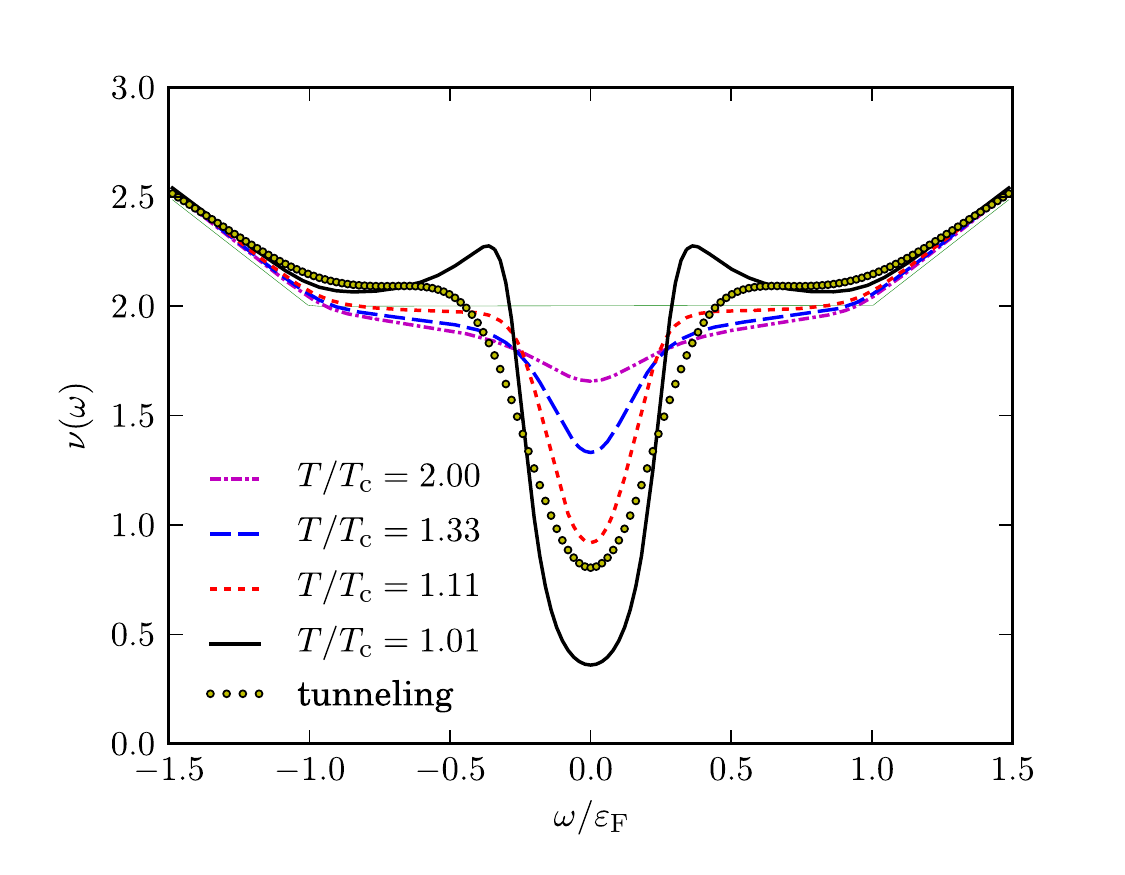}
\begin{center}
\caption{(color online) The density-of-states $\protect\nu(\omega)$ (in units of the non-interacting value $\nu_0$) as a
function of $\omega$ (in units of and measured from the Fermi energy 
$\varepsilon_{\rm F}$). The color coding and labeling is identical
to Fig.~\ref{fig:one}. 
The filled (yellow) circles denote the convolution on the r.h.s. of Eq.~(\ref{AB61}) evaluated at $T/T_{\rm c}= 1.01$.\label{fig:three}}
\end{center}
\end{figure}

The quantity measured in ARPES is the one-body spectral function $\mathcal{A}({\bm k},\omega)$.  
A normal Fermi liquid is characterized by a spectral
function that,  for momenta $|{\bm k}|$ close to the Fermi wave number $k_{\mathrm{F}}$, 
has a single Lorentzian peak as a function of frequency with a
width that vanishes as $(k-k_{\mathrm{F}})^2$~\cite{Giuliani_and_Vignale}. Below we demonstrate that the spectral
function and its integral over ${\bm k}$, \textit{i.e.} the density-of-states (DOS) $\nu(\omega)$, show precursors of the 
DLEC phase of unhybridized MD2DESs at temperatures larger than the
transition temperature. The peak in $\left.\mathcal{A}({\bm k},\omega)\right|_{|{\bm k}|=k_{\mathrm{F}}}$ broadens as one approaches the
transition and splits into two peaks, as can be seen in Fig.~\ref{fig:one}.
At the same time $\nu(\omega)$ develops a strong dip at the Fermi
energy in contrast to the featureless behavior displayed by the non-interacting system (Fig.~\ref{fig:three}).

\textit{MDF Hamiltonian and exciton condensation---}We consider a
double layer hosting MDFs whose top (bottom) layer is p-doped (n-doped). The two layers are coupled only via the Coulomb
interaction between the carriers and electron tunneling is completely
suppressed. Following Refs.~\cite{lozovik_jept_2008,
min_prb_2008,joglekar_prb_2008}, we include in our calculations only the
conduction band (upper Dirac cone) for the n-doped layer and the valence
band (lower Dirac cone) for the p-doped layer. This is justified since the
temperatures we consider are sufficiently small and the screening lengths
sufficiently large, so that the energetically remote bands can, in first instance, be
considered as inert. The band dispersions can be approximated by $\xi_{\mathrm{b%
}}({\bm k}) = v |{\bm k}| - \varepsilon_{\mathrm{F}}$ and $\xi_{\mathrm{t}}({%
\bm k}) = - v |{\bm k}| + \varepsilon_{\mathrm{F}}= - \xi_{\mathrm{b}}({\bm k%
})$ for the bottom (b) and top (t) layers, respectively. Here $v$ is the
Dirac velocity, which is $\approx 10^{6}~\mathrm{m}/\mathrm{s}$ for
double-layer graphene and $\approx 5\times 10^{5}~\mathrm{m}/\mathrm{s}$ for
the surface states of typical known TI thin films~\cite{bandvelo}. We set $\hbar
=1$ throughout this work and include only inter-layer interactions, assuming
that intra-layer interactions can be taken into account by renormalizing the
bare parameters of the two separate-layer Hamiltonians.

We therefore consider the following effective Hamiltonian:
\begin{eqnarray}  \label{eq:Hamiltonian}
{\hat {\mathcal{H}}} &= &\sum_{{\bm k}, i = \mathrm{b},\mathrm{t}} \xi_i({%
\bm k}) {\hat c}^\dagger_{i, {\bm k}} {\hat c}_{i, {\bm k}}  \notag \\
&+& \sum_{{\bm k}_1, {\bm k}_2, {\bm K}} V_0({\bm k}_1, {\bm k}_2, {\bm K}) {%
\hat c}^\dagger_{\mathrm{t}, {\bm k}_1 - {\bm K}} {\hat c}^\dagger_{\mathrm{b%
}, {\bm k}_2} {\hat c}_{\mathrm{b}, {\bm k}_1} {\hat c}_{\mathrm{t}, {\bm k}%
_2-{\bm K}}~,  \notag \\
\end{eqnarray}
where ${\hat c}_{i, {\bm k}}$ (${\hat c}^\dagger_{i, {\bm k}}$) is the
annihilation (creation) operator for an electron in the $i$-th layer with
momentum ${\bm k}$. The bare (repulsive) interaction $V_0({\bm k}_1,{\bm k}%
_2,{\bm K})$ depends on the three wave vectors ${\bm k}_1$, ${\bm k}_2$, and 
${\bm K}$ (and not only on $|{\bm k}_1 - {\bm k}_2|$, which is the momentum
transfer) because of the chirality factors~\cite{reviews},
which encode the suppressed backscattering property of MDFs. After a particle-hole
transformation ${\hat c}_{\mathrm{t}, {\bm k}} \to {\hat h}^\dagger_{\mathrm{%
t}, -{\bm k}}$ in the top layer one immediately sees that the second term in
Eq.~(\ref{eq:Hamiltonian}) describes pairing between an electron with
momentum ${\bm k}_1$ in the bottom layer and a hole with momentum $-{\bm k}%
_1 + {\bm K}$ in the top layer, interacting via an attractive potential
equal to $-V_0({\bm k}_1,{\bm k}_2,{\bm K})$.

\begin{figure}[tbp]
\includegraphics[width = 1.0 \linewidth]{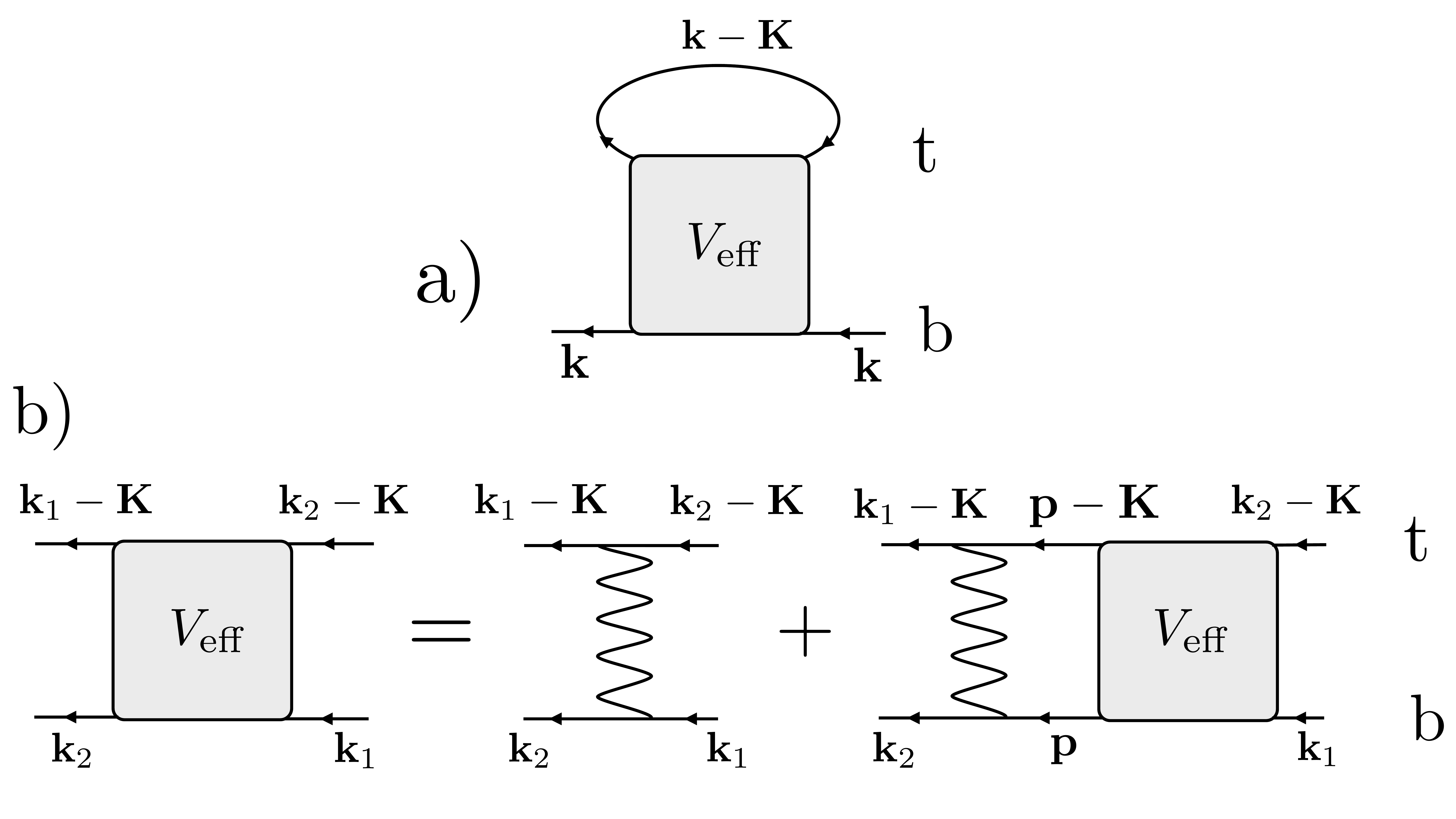}
\begin{center}
\caption{a) The diagram considered for the self energy: the upper line
refers to the top (hole) layer while the lower line to the bottom (electron)
layer. b) The ladder approximation for the effective interaction $V_{\mathrm{%
eff}}$ -- Eq.~(\protect\ref{eq:Vegen}). The thin wavy line is the screened
interlayer interaction $V_0({\bm k}_1, {\bm k}_2, {\bm K})$; ${\bm K}$ is
the center-of-mass momentum and ${\bm p}$ is an integration variable.\label{fig:two}}
\end{center}
\end{figure}

The EC state is signaled by  a pole in the
finite-temperature many-particle scattering amplitude $V_{\mathrm{eff}}$
which appears at a critical temperature $T = T_{\mathrm{c}}$ when the
center-of-mass momentum ${\bm K}$ of a pair and the ``conjugate'' energy $\Omega
= \xi_{\mathrm{t}}({\bm K}-{\bm k}_1) + \xi_{\mathrm{b}}({\bm k}_1)$ vanish.
In Fig.~\ref{fig:two} we show the diagrammatic representation of the
electron self energy [panel a)] and the many-particle scattering amplitude
in the ladder approximation [panel b)]. The ladder sum in panel b) is
analytically intractable for a generic two-body potential $V_0({\bm k}_1, {\bm k}_2, {\bm K})$. To make progress, we approximate~\cite{mink_prl_2012} $V_0({\bm k}_1, {\bm k}_2, {\bm K})$ 
by a momentum-independent constant $U$. (As discussed in Ref.~\cite{mink_prl_2012}, close to $T_{\rm c}$ the chirality factors do not play any
qualitatively significant role.) This approximation is reasonable since it
relies on screening, which is robust when the electron and hole densities in
the two layers are large. Substantial numerical work is needed to transcend
this simple approximation and, in any event, a brute-force numerical
approach would mask the elementary understanding of the physical effect we want to
explain here.

When the bare interaction does not carry momentum labels, the ladder sum
in Fig.~\ref{fig:two} is a geometric series which can be summed to yield 
\begin{equation}  \label{eq:Vegen}
V_{\mathrm{eff}}({\bm K},\Omega_\nu) \equiv \frac{U}{1 - U \Xi({\bm K}%
,\Omega_\nu)}~.
\end{equation}
Here $\Omega_\nu=2\pi \nu/\beta$ are bosonic Matsubara frequencies
with $\beta = (k_{\mathrm{B}} T)^{-1}$, 
\begin{equation}  \label{eq:Pgen}
\Xi({\bm K},\Omega)= \frac{1}{A} \sum_{\bm k} \frac{n_{\rm F}(\xi_{\mathrm{t}}( {\bm k}-{\bm K}) ) - n_{\rm F}(\xi_{\mathrm{b}}({\bm k}))}{\xi_{\mathrm{b}}({\bm k}) -
\xi_{\mathrm{t}}( {\bm k}-{\bm K} ) -\mathrm{i} \Omega}
\end{equation}
is a pairing susceptibility, $A$ is the 2D
electron system area ,and $n_{\rm F}(E) = 1/[1+\exp(\beta E)]$ is the Fermi-Dirac
distribution.  The pairing susceptibility evaluated at ${\bm K}= {\bm 0}$ and 
$\Omega=0$ diverges logarithmically at zero temperature, and therefore the
denominator of Eq.~(\ref{eq:Vegen}) must vanish at some non-zero temperature 
$T_{\mathrm{c}}$ no matter how small the electron-hole attraction. This is
the mean-field pairing transition temperature~\cite{KT}. 
As temperature decreases towards $T_{\mathrm{c}}$, pairing fluctuations first lead to an
enhancement of $V_{\mathrm{eff}}$ for small ${\bm K}$ and $\Omega$ and
ultimately at  $T = T_{\mathrm{c}}$  to a pole at $\Omega = 0$ and ${\bm K} = {\bm 0}$.

\textit{Exciton fluctuations and ARPES spectra. ---} The anomalous
properties of the effective interaction $V_{\rm eff}$ on approaching $T_{\rm c}$ lead to observable consequences~\cite{varlamov}. 
Here we concentrate on the spectral function of the
bottom (n-doped) layer $\mathcal{A}_{\mathrm{b}}({\bm k},\omega)=
-\pi^{-1}\Im m~G^{\mathrm{R}}_{\mathrm{b}}({\bm k},\omega)$, which can be
calculated from the retarded self-energy $\Sigma^{\mathrm{R}}_{\mathrm{b}}({%
\bm k},\omega)$. As is customary~\cite{varlamov}, we proceed by first
evaluating the self-energy on the imaginary-frequency axis and then
analytically continue it to real frequency. According to the diagram shown
in Fig.~\ref{fig:two}a), the self-energy is given by
\begin{equation}  \label{eqn:SelfEnergy}
\Sigma_{\mathrm{b}}({\bm k},\omega_n) = -\frac{1}{\beta A} \sum_{{\bm K}%
,\Omega_\nu} V_{\mathrm{eff}}({\bm K},\Omega_\nu) G^{(0)}_{\mathrm{t}}({\bm k%
}-{\bm K},\omega_n-\Omega_\nu)
\end{equation}
where $\omega_n = 2 \pi (n+1/2)/\beta$ are fermionic Matsubara frequencies
and $G^{(0)}_{\mathrm{t}}({\bm k},\omega_n)=[\mathrm{i} \omega_n-\xi_{%
\mathrm{t}}({\bm k})]^{-1}$ is the free-electron Green's function in the top
layer. In order to calculate the self-energy from Eq.~(\ref{eqn:SelfEnergy})
we approximate $V_{\mathrm{eff}}$ by a form valid~\cite{varlamov} at small ${\bm K}$
and $\Omega$ where fluctuations are strongest:
\begin{eqnarray}  \label{eq:smallKOmega}
V^{\mathrm{R}}_{\mathrm{eff}}({\bm K},\Omega) &\approx & \frac{8/(\pi \beta\nu_0)}{-\mathrm{i} \Omega+\alpha(T)+8\xi_0^2 {\bm K}^2/(\pi \beta) }~,
\end{eqnarray}
where $\alpha(T) = 8\log(T/T_{\mathrm{c}})/(\pi\beta)$ measures the distance
from the critical point, $\xi_0 = [7\zeta(3)v^2 \beta^2]/(32 \pi^2)$ is the
coherence length, and $\nu_0 = \varepsilon_{\mathrm{F}}/(2\pi v^2)$ is the 
single-particle DOS at the MDF Fermi energy. 
Using Eq.~(\ref{eq:smallKOmega}) for the retarded fluctuation propagator
we can then perform the sum over the Matsubara
frequencies in Eq.~(\ref{eqn:SelfEnergy}).  
After
analytic continuation we obtain the following expression for the imaginary
part of the retarded self energy: 
\begin{widetext}
\begin{equation}\label{eqn:selfengreimag}
	\Im m \Sigma^{\rm R}_{\rm b}(\V{k},\omega) = \frac{4}{\pi \beta \nu_0} \; \int \frac{d^2\V{q}}{(2\pi)^2} \frac{\omega+\xi_\rmb(\V{q})}{(\omega+
	\xi_\rmb(\V{q}))^2+\gamma ( \V{k}-\V{q} )^2 } \left [ \tanh \left (\frac{\beta \xi_\rmb(\V{q})}{2}\right )  
	-  \coth \left (\frac{\beta(\omega+ \xi_\rmb(\V{q}))}{2}\right)\right]~, 
\end{equation}
\end{widetext}
where $\gamma({\bm q})= \alpha(T) + 8 \xi_0^2 {\bm q}^2/(\pi \beta)$. 

Once
the imaginary part of the self-energy is known, the real part can be
calculated via the Kramers-Kronig relation~\cite
{Giuliani_and_Vignale,KKnumerically}
\begin{equation}  \label{eqn:KK}
\Re e \Sigma^{\mathrm{R}}_{\mathrm{b}}({\bm k},\omega) = \Re e \Sigma^{%
\mathrm{R}}_{\mathrm{b}}({\bm k},\infty)+\mathcal{P} \int \frac{d \nu}{\pi}%
\frac{\Im m \Sigma^{\mathrm{R}}_{\mathrm{b}}({\bm k},\nu)}{\nu-\omega}~.
\end{equation}
Since $\Re e \Sigma^{\rm R}_{\mathrm{b}}({\bm k},\infty)$ depends 
weakly on momentum ${\bm k}$ near $k_{\mathrm{F}}$ and on temperature 
it can be absorbed by measuring energies relative to the Fermi energy.
In practice we compute the self energy by numerical integration of
Eq.~(\ref{eqn:selfengreimag}) and Eq.~(\ref{eqn:KK}), neglecting $\Re
e\Sigma^{\rm R}_{\mathrm{b}}({\bm k},\infty)$ in the latter. In the weak-coupling
limit we find $\mathcal{A}_{\mathrm{b}}({\bm k},\omega)$ to be a delta
function peaked at some frequency $\Delta \omega$. 
Finally, we calculate $\mathcal{A}_{\mathrm{b}}({\bm k},\omega)$ for various values of $T/T_{\rm c}$ 
by using $\Re e \Sigma^{\rm R}_{\mathrm{b}}({\bm k},\infty) = -\Delta \omega$.
The parameter we change in order to approach the phase transition is thus
the interaction strength $U$ or, equivalenty, $T_{\rm c}$.

\textit{Numerical results---}In Fig.~\ref{fig:one} we plot the sum of the
spectral functions of top and bottom layers, $\mathcal{A}({\bm k},\omega) = 
\mathcal{A}_{\mathrm{b}}({\bm k},\omega) + \mathcal{A}_{\mathrm{t}}({\bm k}%
,\omega)$. Due to the perfect particle-hole symmetry of our model, the
spectral function of the top (hole-doped) layer $\mathcal{A}_{\mathrm{t}}({%
\bm k},\omega)$ is given by $\mathcal{A}_{\mathrm{t}}({\bm k},\omega)=%
\mathcal{A}_{\mathrm{b}}({\bm k},-\omega)$. In Fig.~\ref{fig:one} we
illustrate the dependence of $\mathcal{A}({\bm k},\omega)$ on frequency $%
\omega$ for $|{\bm k}| =k_{\mathrm{F}}$. This plot refers to a fixed
temperature, $T = 0.1~T_{\mathrm{F}}$, and to different values of $T/T_{\rm c}$. We note that the single Lorenzian peak of the spectral
function becomes broader and broader as one approaches the phase transition
and eventually splits into two separate peaks very close to the phase
transition.  Note that the
splitting between the two peaks is very large,
$\sim 5 k_{\mathrm{B}}T_{\mathrm{c}}$ at $T/T_{\rm c} = 1.01$. The two-peak
structure resembles the behavior of the spectral function expected below the
transition temperature.  From
Bogoliubov theory one indeed expects two delta-function peaks at frequencies 
$\Omega_{\pm}=\pm |\Delta|$ where $\Delta$
is the excitation spectrum gap.

From the spectral function one can also calculate the DOS $\nu (\omega )$:
\begin{equation}
\nu (\omega)=\int \frac{d^{2}{\bm k}}{(2\pi )^{2}}\mathcal{A}({\bm k},\omega )~.
\end{equation}
We performed the integral numerically 
after evaluating the integrand for momenta up to $|{\bm k}|=3k_{\mathrm{F}}$. Our
results are shown in Fig.~\ref{fig:three}, where, for
the sake of comparison, we have also plotted the non-interacting result. We
note that as one approaches the phase transition more and more spectral
weight is shifted from the Fermi energy ($\omega =0$) to higher energies.
The resulting dip in the DOS around $\omega =0$ can be seen as a precursor
of the gap in the excitation spectrum that opens up only in the
broken-symmetry phase.

In recent decades the DOS of electronic systems has been measured by using
scanning tunneling microscopy (STM), which probes locally to mitigate the
influence of inhomogeneities and some types of disorder~\cite{sts_review}. 
For example the zero bias anomaly of disordered metals~\cite{AA79} has been successfully explored in this way.
The differential tunneling conductivity is normally well approximated by 
\begin{equation}\label{AB61}
\sigma _{\rm tun}(V) = \frac{dI}{dV} \propto \int_{-\infty }^{\infty }\left[ -\frac{\partial n_{\rm F}(
E) }{\partial E}\right] \nu \left( E+eV\right) dE ~.
\end{equation}
In Eq.~(\ref{AB61}) $I$ and $V$ are respectively the
tunneling current and the bias voltage between the STM tip and the sample. 
When $\nu \left( E \right)$ is smooth on the scale of $k_{\rm B} T$ 
the distribution function on the r.h.s. of Eq.~ (\ref{AB61}) 
acts like a delta function so that $\sigma _{\rm tun}(V) \propto \nu(V)$~\cite{AA79}. 
The situation is very different in the case of STM studies of fluctuating
superconductors above $T_{\rm c}$.  In this case the DOS can have 
structure that is sharp on the scale of the temperature.  The tunneling conductivity 
$\sigma _{\rm tun}(V) $ then has a broad pseudogap structure (with a maximum at $eV_{\rm max}=\pi
k_{\rm B}T_{\rm c}$), resembling the BCS gap below $T_{\rm c}$~\cite{Sacepe10}. This occurs in spite
of the predicted~\cite{ARW70,CCRV90} sharp singularity in the fluctuation contribution to the DOS close to $T_{\rm c}$.
The tunneling conductivity is much smoother as a function of temperature or voltage 
 ($\sigma_{\rm tun}(0)/\sigma_{\rm N} \propto \ln[ T_{\rm c}/(T - T_{\rm c})]$
where $\sigma _{\rm N}$ is normal conductivity of the tunnel junction) than DOS is 
as a function of energy. This distinction ultimately reflects the property that the energetically averaged DOS is 
not altered by order~\cite{VD83}. Similar considerations apply to the EC fluctuation contribution to the tunneling conductivity as
illustrated in Fig.~\ref{fig:three} (circles) where we have plotted the r.h.s. of Eq.~(\ref{AB61}) at $T/T_{\rm c} = 1.01$.

In summary, we have demonstrated that ARPES and STM can be used to probe the
onset of exciton condensation in spatially-separated systems of electrons
and holes created by gating techniques in double-layer graphene and
topological-insulator thin films. The advantage of these experimental probes
is that one does not need to realize independent contacts to two
closely-spaced Dirac layers. These probes can be particularly useful in the
context of topological-insulator thin films where separate contacting is precluded by 
hybridization between top and bottom at the sample edges.  

\textit{Acknowledgments---}Work in Pisa was supported by MIUR through the programs 
``FIRB IDEAS" - Project ESQUI (Grant No. RBID08B3FM) and ``FIRB - Futuro in Ricerca 2010" - Project PLASMOGRAPH (Grant No. RBFR10M5BT), and
by the EU Grants NANO-CTM, IP-SOLID, STREP-QNEMS, and STREP-GEOMDISS. AAV was supported by the FP7-IRSES program, Grant No. 236947 ``SIMTECH".
AHM was supported by SWAN, Welch Foundation grant TBF1473, and by
DOE Division of Materials Sciences and Engineering grant DE-FG03-02ER45958.


\begin{thebibliography}{99}
\bibitem{KKnumerically} 
In Eq.~(\ref{eqn:selfengreimag}) we integrate over
momentum by using values of $|{\bm q}|$ up to $q=10~k_{\mathrm{F}}$.
Performing the Hilbert transform in Eq.~(\ref{eqn:KK}) numerically is quite
tricky and deserves special care. In the present case we achieve convergence
by integrating over the variable $\nu$ in the range $-2\times 10^4 \leq \nu
\leq 2\times10^4$.
\end{thebibliography}
\end{document}